------------------------------------------------------------------------------

\documentstyle[12pt,a4]{article}

\begin{document}

\newcommand{\pb}{\bar P}
\newcommand{\pbp}{\bar P'}
\newcommand{\delp}{\Delta P}
\newcommand{\gf}{G_{\mbox{{\scriptsize F}}}}
\newcommand{\order}{{\cal O}}
\newcommand{\leqsim}{\,\mbox{{\scriptsize $\stackrel{<}{\sim}$}}\,}
\newcommand{\heff}{{\cal H}_{\mbox{{\scriptsize eff}}}(\Delta B=-1)}
\newcommand{\heffp}{{\cal H}_{\mbox{{\scriptsize eff}}}^{\mbox{{\scriptsize
pen}}}(\Delta B=-1)}
\newcommand{\VmA}{\mbox{{\scriptsize V--A}}}
\newcommand{\VpA}{\mbox{{\scriptsize V+A}}}
\newcommand{\VpmA}{\mbox{{\scriptsize V$\pm$A}}}
\newcommand{\beq}{\begin{equation}}
\newcommand{\eeq}{\end{equation}}
\newcommand{\bea}{\begin{eqnarray}}
\newcommand{\eea}{\end{eqnarray}}
\newcommand{\non}{\nonumber}
\newcommand{\lab}{\label}
\newcommand{\la}{\langle}
\newcommand{\ra}{\rangle}
\newcommand{\np}{Nucl.\ Phys.}
\newcommand{\pl}{Phys.\ Lett.}
\newcommand{\prl}{Phys.\ Rev.\ Lett.}
\newcommand{\pr}{Phys.\ Rev.}
\newcommand{\zp}{Z.\ Phys.}

\setcounter{page}{0}
\thispagestyle{empty}
\begin{flushright}
MPI-PhT/94-56\\
TUM-T31-69/94\\
August 1994
\end{flushright}
\vspace*{0.2cm}
\begin{center}
{\Large{\bf Limitations in Measuring the Angle $\beta$}}\\
\vspace{0.4cm}
{\Large{\bf by using $SU(3)$ Relations for $B$-Meson}}\\
\vspace{0.4cm}
{\Large{\bf Decay-Amplitudes \footnote[1]{Supported by
the German Bundesministerium f\"ur
Forschung und Technologie under contract 06 TM 732 and by the CEC
science project SC1--CT91--0729.}}}\\
\vspace{1.3cm}
{\large{\sc Andrzej J. Buras}} $^{1,2}$\quad {\large and} \quad
{\large{\sc Robert Fleischer}} ${^1}$\\
\vspace{1cm}
{\sl $^1$ Technische Universit\"at M\"unchen, Physik Department\\
D-85748 Garching, Germany}\\
\vspace{0.5cm}
{\sl $^2$ Max-Planck-Institut f\"ur Physik\\
-- Werner-Heisenberg-Institut --\\
F\"ohringer Ring 6, D-80805 M\"unchen, Germany}\\
\vspace{1cm}
{\large{\bf Abstract}}\\
\vspace{0.8cm}
\end{center}

Flavour $SU(3)$ symmetry of strong interactions and
certain dynamical assumptions have been used in a series of
recent publications to extract weak CKM phases from $B$-decays
into $\{\pi\pi,\pi K, K\bar K\}$ final states. We point out that
irrespectively of $SU(3)$-breaking effects the presence of
QCD-penguin contributions with internal $u$- and $c$-quarks precludes
a clean determination of the angle $\beta$ in the unitarity triangle
by using the branching
ratios only. This difficulty can be overcome by measuring in addition
the ratio $x_d/x_s$ of $B^0_d-\bar B^0_d$ to $B^0_s-\bar B^0_s$
mixings. The measurement of the angle $\gamma$ is unaffected by these
new contributions. Some specific uncertainties related to
$SU(3)$-breaking effects and electroweak penguin contributions
are briefly discussed.

\newpage

Recently in a series of interesting publications \cite{grl}-\cite{lon},
$SU(3)$ flavour symmetry of strong
interactions \cite{zep}-\cite{hmt} has been combined with certain
dynamical assumptions (neglect of annihilation diagrams, etc.) to derive
simple relations among $B$-decay amplitudes into $\pi\pi$, $\pi K$
and $K\bar K$ final states. These $SU(3)$ relations should
allow to determine in a clean manner
both weak phases of the Cabibbo-Kobayashi-Maskawa-matrix (CKM-matrix)
\cite{km} and strong final state interaction phases by measuring {\it
only} branching ratios of the relevant $B$-decays. Neither tagging nor
time-dependent measurements are needed!

In this note we would like to point out certain limitations of this
approach. Irrespectively of the uncertainties related to
$SU(3)$-breaking effects, which have been partially addressed in
\cite{grl}-\cite{lon}, the success of this approach depends on whether
the penguin amplitudes are fully dominated by the diagrams with
internal top-quark exchanges.
As we will show below, sizable contributions may also arise from
QCD-penguins with internal up- and charm-quarks. The main purpose of
our letter is to analyze the impact of these new contributions on the
analyses of refs.~\cite{grl}-\cite{lon}.

Interestingly enough the determination of the angle $\gamma$ in the
unitarity triangle as outlined in \cite{grl,hl,lon} is not affected by
the presence of QCD-penguins with internal $u$- and $c$-quarks.
Unfortunately these new contributions preclude a clean determination
of the angle $\beta$ by using the branching ratios only. We show
however that the additional knowledge of the ratio $x_d/x_s$ of
$B^0_d-\bar B^0_d$ to $B^0_s-\bar B^0_s$ mixings would allow a clean
determination of $\beta$ except for $SU(3)$-breaking uncertainties.

In order to discuss these effects, let us denote, as in
\cite{grl}-\cite{lon}, the amplitudes corresponding to $b\to d$ and $b\to s$
QCD-penguins by $\pb$ and $\pbp$, respectively, and those representing
the CP-conjugate processes by $P$ and $P'$ (these amplitudes can be
obtained easily from $\pb$ and $\pbp$ by changing the signs of the
weak CKM-phases). Then, taking into account
QCD-penguin diagrams with internal $u$-, $c$- and $t$-quarks, we get
\beq\lab{e1}
\begin{array}{rcl}
\pb&=&\sum\limits_{q=u,c,t}V_{qd}^\ast V_{qb}P_q=v_c^{(d)}(P_c-P_u)+v_t^{(d)}
(P_t-P_u)\\
\pbp&=&\sum\limits_{q=u,c,t}V_{qs}^\ast
V_{qb}P_q=v_c^{(s)}(P_c'-P_u')+v_t^{(s)}
(P_t'-P_u'),
\end{array}
\eeq
where we have employed unitarity of the CKM-Matrix and have defined
the CKM-factors as
\beq\lab{e2}
\begin{array}{rcl}
v_c^{(q)}&=&V_{cq}^\ast V_{cb}\nonumber\\
v_t^{(q)}&=&V_{tq}^\ast V_{tb}.
\end{array}
\eeq
Applying the Wolfenstein parametrization \cite{wolf} gives
\beq\lab{e3}
\begin{array}{rcl}
v_c^{(d)}&=&-\lambda|V_{cb}|\left(1+\order(\lambda^4)\right)\\
v_t^{(d)}&=&|V_{td}|\exp{(i\beta)}
\end{array}
\eeq
and
\beq\lab{e4}
\begin{array}{rcl}
v_c^{(s)}&=&|V_{cb}|\left(1+\order(\lambda^2)\right)\\
v_t^{(s)}&=&-|V_{cb}|\left(1+\order(\lambda^2)\right),
\end{array}
\eeq
where the estimate of non-leading terms follows ref.~\cite{blo}. In
order to simplify the presentation we will omitt these non-leading
terms in $\lambda$ in our analysis.

Introducing the notation
\beq\lab{e5}
P_{q_1q_2}\equiv\left|P_{q_1q_2}\right|\exp{(i\delta_{q_1q_2})}\equiv
P_{q_1}-P_{q_2}
\eeq
with $q_1, q_2\in\{u,c,t\}$ and combining eqs.~(\ref{e3}) and (\ref{e4})
with (\ref{e1}) yields
\bea
\pb&=&\left[-\frac{1}{R_t}\frac{|P_{cu}|e^{i\delta_{cu}}}
{|P_{tu}|e^{i\delta_{tu}}}+e^{i\beta}\right]|V_{td}||P_{tu}|
e^{i\delta_{tu}}\label{e6a}\\
\pbp&=&\left[-\frac{|P_{cu}'|e^{i\delta_{cu}'}}
{|P_{tu}'|e^{i\delta_{tu}'}}+1\right]e^{i\pi}|V_{cb}||P_{tu}'|
e^{i\delta_{tu}'}.\lab{e6b}
\eea
$R_t$ is given by the CKM-combination
\beq\lab{e7}
R_t\equiv\frac{1}{\lambda}\frac{|V_{td}|}{|V_{cb}|}
\eeq
and represents the side of the so-called unitarity triangle that is
related to $B^0_d$--$\bar B^0_d$ mixing.  From present
experimental data, we expect $R_t$ being of $\order(1)$ \cite{blo}.

Assuming $SU(3)$ flavour symmetry of strong interactions, the
``primed'' amplitudes $|P_{q_1q_2}'|$ and strong phase shifts
$\delta_{q_1q_2}'$ are equal to the ``unprimed'' ones
\cite{hlgr}-\cite{lon}.
Consequently, the penguin-amplitudes (\ref{e6a}) and (\ref{e6b}) can
be expressed in the form
\bea
\pb&=&\left[-\frac{1}{R_t}\delp +e^{i\beta}\right]|V_{td}||P_{tu}|
e^{i\delta_{tu}}\lab{e9a}\\
\pbp&=&\left[-\delp +1\right]e^{i\pi}|V_{cb}||P_{tu}|
e^{i\delta_{tu}},\lab{e9b}
\eea
where $\delp$ is defined by
\beq\lab{e10}
\delp\equiv|\delp|e^{i\delta_{\delp}}\equiv\frac{|P_{cu}|e^{i\delta_{cu}}}
{|P_{tu}|e^{i\delta_{tu}}}
\eeq
and describes the contributions of the QCD-penguins with internal
$u$- and $c$-quarks. Notice that $\delp$ suffers from large hadronic
uncertainties, in particular
from strong final state interaction phases parametrized by
$\delta_{cu}$ and $\delta_{tu}$. In the limit of degenerate $u$- and
$c$-quark masses, $\delp$ would vanish due to the GIM mechanism.
However, since $m_u\approx4.5$~MeV, whereas $m_c\approx1.3$~GeV, this GIM
cancellation is incomplete and in principle sizable effects arising
from $\delp$ could be expected.

In order to investigate this issue quantitatively, let us estimate
$\delp$ by using the perturbative approach of
Bander, Silverman and Soni~\cite{bss}.
To simplify the following discussion, we
neglect the influence of the renormalization group evolution from
$\mu=\order(M_W)$ down to $\mu=\order(m_b)$ and take into account QCD
renormalization effects only approximately through the replacement
$\alpha_s\to \alpha_s(\mu)$. Then, the low-energy effective
penguin Hamiltonian is given by (see, e.g., refs.~\cite{hw}-\cite{rf})
\bea
\heffp&=&-\frac{\gf}{\sqrt{2}}\frac{\alpha_s(\mu)}
{8\pi}\sum\limits_{q=d,s}
\biggl[v_c^{(q)}\left\{G(m_c,k,\mu)-G(m_u,k,\mu)\right\}\lab{e11}\\
&&+v_t^{(q)}\left\{E(x_t)+\frac{2}{3}\ln\left(\frac{\mu^2}{M_W^2}\right)
-G(m_u,k,\mu)
\right\}\biggr]P^{(q)},\nonumber
\eea
where
\beq\lab{e12}
P^{(q)}=-\frac{1}{3}Q_3^{(q)}+Q_4^{(q)}-\frac{1}{3}Q_5^{(q)}+Q_6^{(q)}
\eeq
is a linear combination of the usual QCD-penguin operators
\beq\lab{e13}
\begin{array}{rcl}
Q_{3}^{(q)}&=&(\bar qb)_{\VmA}\sum_{q'}(\bar q'q')_{\VmA}\\
Q_{4}^{(q)}&=&(\bar q_{\alpha}b_{\beta})_{\VmA}\sum_{q'}
(\bar q'_{\beta}q'_{\alpha})_{\VmA}\\
Q_{5}^{(q)}&=&(\bar qb)_{\VmA}\sum_{q'}(\bar q'q')_{\VpA}\\
Q_{6}^{(q)}&=&(\bar q_{\alpha}b_{\beta})_{\VmA}\sum_{q'}
(\bar q'_{\beta}q'_{\alpha})_{\VpA}
\end{array}
\eeq
and the function $G(m,k,M)$ is defined by \cite{rf}
\beq\lab{e14}
G(m,k,M)\equiv-4\int\limits_0^1dxx(1-x)\ln\left[\frac{m^2-k^2x(1-x)}
{M^2}\right].
\eeq
The four-vector $k$ denotes the momentum of the virtual gluon
appearing in the QCD-penguin diagrams, $x_t=m_t^2/M_W^2$ and
\beq
E(x)=-\frac{2}{3}\ln x+\frac{x^2(15-16x+4x^2)}{6(1-x)^4}\ln x +
\frac{(18-11x-x^2)x}{12(1-x)^3}
\eeq
is one of the so-called Inami-Lim functions~\cite{il}. In
eq.~(\ref{e13}), $q'$ runs over the quark flavours being active at the
scale $\mu=\order(m_b)$ $(q'\in\{u,d,c,s,b\})$ and $\alpha$, $\beta$ are
$SU(3)_{\mbox{{\scriptsize C}}}$ colour indices.

Evaluating hadronic
matrix elements of $\heffp$ and comparing them with eq.~(\ref{e1}), we find
\beq\lab{e15}
\delp\approx\frac{G(m_c,k,\mu)-G(m_u,k,\mu)}{E(x_t)+\frac{2}{3}\ln\left(
\frac{\mu^2}{M_W^2}\right)-G(m_u,k,\mu)}.
\eeq
In this perturbative approximation, the strong phase shift of $\delp$
is generated exclusively through absorptive parts of the penguin
amplitudes with internal $u$- and $c$-quarks
(``Bander--Silverman--Soni mechanism''~\cite{bss}).
Whereas the $\mu$-dependence
cancels exactly in (\ref{e15}), $\delp$ depends strongly on the value of
$k^2$, as can be seen from Figs.~1 and 2. Simple kinematical
considerations at the quark-level imply that $k^2$ should lie within
the ``physical'' range~\cite{gh,rf}
\beq\lab{e16}
\frac{1}{4}\leqsim\frac{k^2}{m_b^2}\leqsim\frac{1}{2}.
\eeq
For such values of $k^2$, we read off from Figs.~1 and 2 that
\beq\lab{e16a}
0.2\leqsim|\delp|\leqsim0.5\quad\mbox{and}\quad70^\circ\leqsim
\delta_{\delp}\leqsim130^\circ,
\eeq
respectively. Consequently, $\delp$ may lead to sizable effects in the
$SU(3)$ triangle relations discussed below. We are aware of the fact
that the estimate of $\delp$ given here is very rough. It illustrates
however a potential hadronic uncertainty which cannot be ignored.

In refs.~\cite{grl}-\cite{lon}, only QCD-penguins with internal top-quarks
have been taken into account. This approximation corresponds to
$\delp=0$ and gives
\bea
\pb_{\delp=0}&=&a_Pe^{i\beta}e^{i\delta_P}\lab{e17a}\\
\pbp_{\delp=0}&=&a_{P'}e^{i\pi}e^{i\delta_P},\lab{e17b}
\eea
where
\beq\lab{e18}
a_P=|V_{td}||P_{tu}|,\quad a_{P'}=a_P/(\lambda R_t)\quad\mbox{and}\quad
\delta_P=\delta_{tu}.
\eeq
Notice that the weak- and strong phase structure of (\ref{e17b}) is
similar to (\ref{e9b}) which can be re-written in the form
\beq\lab{e19}
\pbp=\rho_{P'}a_{P'}e^{i\pi}e^{i(\delta_P-\psi')}
\eeq
with
\beq\lab{e20a}
\rho_{P'}=\sqrt{1-2|\delp|\cos\delta_{\delp}+|\delp|^2}
\eeq
and
\beq\lab{e20b}
\tan\psi'=\frac{|\delp|\sin\delta_{\delp}}{1-|\delp|\cos\delta_{\delp}}.
\eeq
In eq.~(\ref{e19}), $\pi$ represents the CP-violating weak phase, while
$\delta_P-\psi'$ denotes the CP-conserving strong phase shift.

Therefore, the determination of the weak CKM-angle $\gamma$ through
$SU(3)$ triangle relations involving the charged $B$-meson decays
$B^+\to \{\pi^0K^+,\pi^+K^0,\pi^+\pi^0\}$ (and the corresponding
CP-conjugate modes) as outlined in refs.~\cite{grl,hl,lon} is not affected by
$\delp$ at all, since no non-trivial weak phases appear in $P'$
$(\pbp)$ even in the presence of QCD penguins with internal $u$- and
$c$-quarks. However, the strong phase differences
$\delta_P-\delta_{T,C}$ are shifted by the angle $\psi'$.
Here $\delta_T$ and $\delta_C$ denote the
strong phases of the ``tree'' and ``colour-suppressed''  amplitudes
\beq\lab{e21a}
T=a_Te^{i\gamma}e^{i\delta_T}\quad\mbox{and}\quad
C=a_Ce^{i\gamma}e^{i\delta_C}
\eeq
contributing to $B^\pm\to \pi^\pm\pi^0$, respectively.

On the other hand, the QCD-penguin contributions with internal $u$-
and $c$-quarks affect the extraction of the phase
$\beta$ by using the triangle relations~\cite{hlgr}-\cite{lon}
\beq\lab{e21}
\begin{array}{rcl}
A(B^0_d\to\pi^+\pi^-)+\sqrt{2}A(B^0_d\to\pi^0\pi^0)&=&\sqrt{2}
A(B^+\to\pi^+\pi^0)\\
(T+P)\qquad+\qquad(C-P)&=&(T+C)
\end{array}
\eeq
and
\beq\lab{e22}
\begin{array}{rcl}
A(B^0_d\to\pi^- K^+)/r_u+
\sqrt{2}A(B^0_d\to\pi^0 K^0)/r_u&=&\sqrt{2}
A(B^+\to\pi^+\pi^0)\\
(T+P'/r_u)\qquad+\qquad(C-P'/r_u)
&=&(T+C),
\end{array}
\eeq
where $r_u=V_{us}/V_{ud}$.

Following the approach outlined in ref.~\cite{lon}, the complex
amplitudes $P'$ and $P$ can be determined up to a {\it common} strong phase
shift (and some discrete ambiguities) through a two-triangle
construction involving the rates of the five modes appearing in
(\ref{e21}) and (\ref{e22}) and two additional rates that determine
$|P|$ and $|P'|$
(e.g., $B^+\to K^+ \bar K^0$ and $B^+\to\pi^+ K^0$, respectively).
Therefore, the relative angle $\vartheta$ between $P$ and $P'$
can be measured. Expressing $P$ in the form
\beq\lab{e23}
P=\rho_Pa_Pe^{-i\beta}e^{i(\delta_P-\psi)}
\eeq
with
\beq\lab{e24a}
\rho_{P}=\frac{1}{R_t}\sqrt{R_t^2-2R_t|\delp|\cos(\beta+\delta_{\delp})+
|\delp|^2}
\eeq
and
\beq\lab{e24b}
\tan\psi=\frac{|\delp|\sin(\beta+\delta_{\delp})}
{R_t-|\delp|\cos(\beta+\delta_{\delp})},
\eeq
we find using (\ref{e18}), (\ref{e19}) and (\ref{e23})
\beq\lab{e25a}
\frac{1}{r_t}\frac{P'}{P}=\frac{\rho_{P'}}{\rho_P}e^{i(\psi-\psi')}
\equiv\frac{\rho_{P'}}{\rho_P}e^{i(\vartheta-\beta)},
\eeq
where $r_t\equiv V_{ts}/V_{td}$. Note that the deviation of the rhs.\
of eq.~(\ref{e25a}) from one represents corrections to the relation
between $P'$ and $P$ presented in refs.~\cite{ghlr}-\cite{lon}.
Consequently, $\vartheta$ is given by
\beq\lab{e25}
\vartheta=\beta+\psi-\psi'.
\eeq
In contrast to $\psi'$, which is a pure strong phase, $\psi$ is a
combination of both CP-conserving strong phases $(\delta_{\delp})$ and the
CP-violating weak phase $\beta$.

If we neglect the QCD-penguins with internal $u$- and $c$-quarks, as
the authors of refs.~\cite{hlgr}-\cite{lon}, we have $\delp=0$
and, thus, $\vartheta$ is
equal to the CKM-angle $\beta$ in this approximation. However, as can
be seen from Figs.~1 and 2, the perturbative estimates of $\delp$
indicate that sizable contributions may arise from this amplitude
which show up in eq.~(\ref{e25}) as the phase difference $\psi-\psi'$.
Since both $\psi$ and $\psi'$ contain strong phases,
$\vartheta$ is not a theoretical clean quantity in general (even if the
$SU(3)$ triangle relations were valid exactly!) and this determination
of the angle $\beta$ suffers from hadronic uncertainties in contrast
to the assertions made in \cite{hlgr}-\cite{lon}.

In order to illustrate this point quantitatively,
we have plotted the dependence of
$\psi-\psi'$ on $k^2/m_b^2$ arising from (\ref{e15}) for $R_t=1$ and
various angles $\beta$ in Fig.~3. The corresponding curves for
$\rho_{P'}/\rho_P$ (see eq.~(\ref{e25a})) are shown in Fig.~4. In
drawing these figures, we have taken into account that the angle
$\beta$ is smaller than $45^\circ$ for the present range of
$|V_{ub}/V_{cb}|$~\cite{blo}. Notice that the hadronic
uncertainties in (\ref{e25a}) and (\ref{e25}) cancel each other,
i.e., $P'=r_tP$ and $\psi'=\psi$, if we choose
$R_t=1$ and $\beta=0$. This cancellation is, however,
incomplete in the general case.

As an illustration consider a measurement of $\vartheta=15^\circ$.
Setting $\delp=0$ one would conclude that $\beta=15^\circ$ and
$\sin2\beta=0.50$. With $\delp\not=0$, as calculated here, the true
$\beta$ could be as high as $20^\circ$ $(\psi-\psi'=-5^\circ)$ giving
$\sin2\beta=0.64$. We observe that this uncertainty (in addition to
possible $SU(3)$-breaking effects) could spoil the comparison of
$\beta$, measured this way, with the clean determination of
$\sin2\beta$ in $B_d\to\psi K_{\mbox{{\scriptsize S}}}$.

We now want to demonstrate that the hadronic uncertainties affecting
the determination of $\beta$ through (\ref{e25}) can be eliminated
provided $R_t$ is known. To this end, we consider the
``normalized'' penguin amplitudes
\bea
\frac{1}{\lambda|V_{cb}|}P&=&\left[-\delp+R_t
e^{-i\beta}\right]|P_{tu}|e^{i\delta_{tu}}\lab{e26a}\\
\frac{1}{|V_{cb}|}P'&=&\left[\delp-1\right]|P_{tu}|e^{i\delta_{tu}}\lab{e26b}
\eea
and those of the corresponding CP-conjugate processes
(see (\ref{e9a}) and (\ref{e9b})) which are related
to (\ref{e26a}) and (\ref{e26b}) through the substitution $\beta\to-\beta$.
Combining these complex amplitudes in the form
\beq\lab{e27}
z\equiv\frac{P+\lambda
P'}{\pb+\lambda\pbp}=\frac{1-R_te^{-i\beta}}{1-R_te^{i\beta}}=e^{i2\gamma},
\eeq
we observe that both $\delp$ and $|P_{tu}|\exp(i\delta_{tu})$, which
are unknown, non-perturbative quantities, cancel
in the ratio $z$. The appearance of $\gamma$ in this ratio can be
understood by noting that
\beq\lab{e27c}
\bar P+\lambda\bar
P'=-v_u^{(d)}P_{tu}=-|V_{ub}|e^{-i\gamma}(1+\order(\lambda^2))
|P_{tu}|e^{i\delta_{tu}}.
\eeq
Consequently, in the limit of exact $SU(3)$
triangle relations (\ref{e21}) and (\ref{e22}), the angle $2\gamma$,
which is related
to $\beta$ through
\beq\lab{e28}
\tan2\gamma=\frac{2R_t(1-R_t\cos\beta)\sin\beta}
{1-2R_t\cos\beta+R_t^2\cos2\beta},
\eeq
can be also here extracted without theoretical uncertainties. If, in
addition, $R_t$ is also known, the CKM-phase $\beta$ can be determined
as well. In Fig.~5, we have illustrated the dependence of $2\gamma$ on
$\beta$ for various values of $R_t$. Note that
$2\gamma=\pi-\beta$, if $R_t=1$.

The theoretically cleanest way of measuring $R_t$ without using
CP-violating quantities is obtained through
\beq\lab{e29}
R_t=\frac{1}{\sqrt{R_{ds}}}\sqrt{\frac{x_d}{x_s}}\frac{1}{|V_{us}|},
\eeq
where $x_d$ and $x_s$ give the sizes of $B^0_d-\bar B^0_d$ and
$B^0_s-\bar B^0_s$ mixings, respectively, and
\beq\lab{e30}
R_{ds}=\frac{\tau_{B_d}}{\tau_{B_s}}\cdot\frac{m_{B_d}}{m_{B_s}}
\left[\frac{F_{B_d}\sqrt{B_{B_d}}}{F_{B_s}\sqrt{B_{B_s}}}\right]^2
\eeq
summarizes the $SU(3)$ flavour-breaking effects. In the strict
$SU(3)$ limit, we have $R_{ds}=1$. The main theoretical uncertainty
resides in the values of the $B$-meson decay constants $F_{B_{d,s}}$
and in the non-perturbative parameters $B_{B_{d,s}}$ which parametrize
the hadronic matrix elements of the relevant operators. We believe
however that $R_{ds}$ can be more reliably estimated than $\delp$.

At this point, it should be stressed that the elimination of the
hadronic uncertainties arising from $\delp$, i.e., the QCD-penguins
with internal $u$- and $c$-quarks, requires to consider also the
CP-conjugate modes to extract ``clean'' values of $\beta$.
Furthermore, $R_t$ has to be known. These
complications are very different from the situation in
refs.~\cite{hlgr}-\cite{lon}, where it has been emphasized that it was
not necessary to measure the charge-conjugate rates in order to
determine $\beta$.

Assuming {\it factorization}, $SU(3)$-breaking corrections can be
taken into account approximately through the substitutions $r_u\to r_u
f_K/f_\pi$ \cite{grl}-\cite{lon} and $r_t\to r_tf_K/f_\pi$ in
eqs.~(\ref{e22}) and (\ref{e25a}), respectively, where $P'$ and $P$ in
eq.~(\ref{e25a}) are the same as in the triangle relations
(\ref{e21}) and (\ref{e22}). Moreover, we have to
replace $\lambda$ in our result (\ref{e27}) by $\lambda f_\pi/f_K$.
$SU(3)$-breaking effects must also be taken into account in the
determination of $|P|$ and $|P'|$ from
the decay amplitudes $|A(B^+\to K^+\bar K^0)|$ and
$|A(B^+\to\pi^+K^0)|$, respectively. Within the framework of
factorization we find
\bea
|P|&=&\frac{f_\pi}{f_K}\frac{F_{B\pi}(0;0^+)}{F_{BK}(0;0^+)}|A(B^+\to
K^+ \bar K^0)|\lab{e31b}\\
|P'|&=&|A(B^+\to\pi^+K^0)|\lab{e31a},
\eea
where $F_{B\pi}(0;0^+)$ and $F_{BK}(0;0^+)$ are form factors
parametrizing the hadronic quark-current matrix elements $\langle\pi^+
|(\bar bd)_{\VmA}|B^+\rangle$ and $\langle K^+|(\bar
bs)_{\VmA}|B^+\rangle$, respectively \cite{bgr}.
Unfortunately, hadronic form factors appear in eq.~(\ref{e31b}) which
are model dependent. Using, for example, the model of Bauer, Stech and
Wirbel \cite{bsw}, we estimate that the $SU(3)$-breaking factor in
(\ref{e31b}) should be of $\order(0.7)$.

At present, there is no reliable theoretical technique available to evaluate
non-factorizable $SU(3)$-breaking corrections to the relevant
$B$-decays. Since already the factorizable
corrections are quite large $((20-30)\%)$, we expect that
non-factorizable $SU(3)$-breaking may also lead to sizable effects. In
particular, such corrections could spoil the elimination of the
QCD-penguins with internal $u$- and $c$-quarks through eq.~(\ref{e27}).
Furthermore, in the presence of a heavy top-quark, electroweak-penguin
contributions may also lead to sizable corrections ($(10-30)\%$ at the
amplitude level) to the penguin sectors of $B$-decays into final
states that contain mesons with a CP-self-conjugate quark
content \cite{dun}-\cite{dh}. Possible impact of electroweak penguins on the
approach of refs.~\cite{grl}-\cite{lon} has been recently also
emphasized in ref.~\cite{dh2}.

In summary, we have shown that QCD-penguins with internal $u$- and
$c$-quarks may lead to sizable systematic errors in the
extraction of the CKM-phase $\beta$ by using the approach presented in
refs.~\cite{hlgr}-\cite{lon}. However, $\beta$ can still be determined
in a theoretical clean way (up to corrections arising from
non-factorizable $SU(3)$-breaking and certain neglected contributions which
are expected to be small on dynamical grounds \cite{grl}-\cite{lon}), if
$R_t$ and the rates of the CP-conjugate processes appearing in the
corresponding triangle relations are measured. On the other hand, the
determination of $\gamma$ along the lines suggested in
\cite{grl}-\cite{lon} and in (\ref{e27}) in the present paper is not
affected by these new QCD-penguin contributions. Its fate depends then
only on the ability of estimating $SU(3)$-breaking effects and on the
precision with which the relevant branching ratios can be measured one
day.

\newpage

\section*{Figure Captions}
\begin{table}[h]
\begin{tabular}{ll}
Fig.\ 1:&The dependence of $|\delp|$ on $k^2/m_b^2$.\\
&\\
Fig.\ 2:&The dependence of $\delta_{\delp}$ on $k^2/m_b^2$.\\
&\\
Fig.\ 3:&The dependence of $\psi-\psi'$ on $k^2/m_b^2$
for $R_t=1$ and various values of\\
&the CKM-angle $\beta$.\\
&\\
Fig.\ 4:&The dependence of $\rho_{P'}/\rho_P$ on $k^2/m_b^2$
for $R_t=1$ and various values of\\
&the CKM-angle $\beta$.\\
&\\
Fig.\ 5:&The dependence of angle $2\gamma$ on the CKM-angle $\beta$
for various values\\
&of $R_t$.\\
\end{tabular}
\end{table}

\end{document}